\documentclass[runningheads]{llncs}
\usepackage[T1]{fontenc}
\renewcommand{\texttt}[1]{\textnormal{\ttfamily\hyphenchar\font=`\-#1}}
\usepackage{array}
\usepackage{geometry}
\usepackage{booktabs}
\usepackage{graphicx}
\usepackage{mdframed}
\usepackage{soul}
\usepackage{xcolor}
\begin{document}
\title{Synthline: A Product Line Approach for Synthetic Requirements Engineering Data Generation using Large Language Models}
\titlerunning{Synthline}
%
\author{Abdelkarim El-Hajjami\orcidID{0009-0004-7053-3264} \and \\
Camille Salinesi\orcidID{0000-0002-1957-0519}}
\authorrunning{A. El-Hajjami and C. Salinesi}
%
\institute{Paris 1 Panthéon–Sorbonne University, Paris, France\\
\email{\{abdelkarim.el-hajjami, camille.salinesi\}@univ-paris1.fr}}

\maketitle              
\begin{abstract}
While modern Requirements Engineering (RE) heavily relies on natural language processing and Machine Learning (ML) techniques, their effectiveness is limited by the scarcity of high-quality datasets. This paper introduces Synthline, a Product Line (PL) approach that leverages
Large Language Models to systematically generate synthetic RE data for classification-based use cases. Through an empirical evaluation conducted in the context of using ML for the identification of requirements specification defects, we investigated both the diversity of the generated data and its utility for training downstream models. Our analysis reveals that while synthetic datasets exhibit less diversity than real data, they are good enough to serve as viable training resources. Moreover, our evaluation shows that combining synthetic and real data leads to substantial performance improvements. Specifically, hybrid approaches achieve up to 85\% improvement in precision and a 2× increase in recall compared to models trained exclusively on real data. These findings demonstrate the potential of PL-based synthetic data generation to address data scarcity in RE. We make both our implementation and generated datasets publicly available to support reproducibility and advancement in the field.
 \and Product Line
\end{abstract}
\section{Introduction}
Requirements Engineering (RE) largely operates through Natural Language (NL), which is the primary medium for capturing, specifying, and validating all types of requirements \cite{b1}. Although this reliance on NL has enabled significant advancements in the use of Natural Language Processing (NLP) techniques and Artificial Intelligence (AI) models in RE \cite{b11,b12}, their effectiveness remains tied to the availability of diverse high-quality datasets. Even with the paradigm shift introduced by in-context learning in Large Language Models (LLMs) \cite{b9}, achieving optimal performance still requires sufficient high-quality training data and additional fine-tuning \cite{b10}. Therefore, the shortage of freely available RE-specific datasets remains a major bottleneck \cite{b11},\cite{b24}, slowing progress and limiting the effective use of AI models in RE.

However, the acquisition of high-quality data presents significant challenges. It is both costly and time-consuming, with the data annotation phase often being laborious and error-prone \cite{b15}. In addition, companies are often reluctant to share sensitive and private data, restricting the availability and use of such data for research purposes outside the organizations that produce it. Consequently, synthetic data emerges as a promising solution to address these issues.

Data synthesis may take several forms: anonymization of sensitive data \cite{b22}; annotation for supervised learning \cite{b15}; augmentation through text transformations (e.g., word deletion or synonym replacement) \cite{b6,b8}; and the generation of entirely new samples from scratch \cite{b34}. Our research follows this last approach, leveraging LLMs’ advanced text generation and multilingual capabilities \cite{b2} to address data scarcity in RE. This approach not only enables cost-effective, large-scale data generation but also provides systematic control over data properties, supporting varied RE use cases and addressing specific challenges such as data imbalance \cite{b21} and the scarcity of non-English resources \cite{b11}.

To enable systematic controllability, we propose a generative approach that formalizes the variability of RE use cases and their specific requirements using a Feature Model (FM). This study represents the first iteration of a broader design science cycle \cite{b3}, where we design and evaluate a minimal Product Line (PL) of synthetic datasets designed for classification-based RE use cases.
Within this scope, we examine the following research questions:
\begin{itemize}
    \item \textbf{RQ1.} How does the diversity of the synthetic data compare to that of real data?
    \item \textbf{RQ2.} How does training on synthetic data affect model performance relative to training exclusively on real data?
\end{itemize}
\textbf{Contributions.}
\textcircled{\raisebox{-0.8pt}{1}} We propose \textit{Synthline}, a minimal and configurable approach for systematically controlling LLM-based data generation.
\textcircled{\raisebox{-0.8pt}{2}} We evaluate our approach on the use case of identifying requirements specification defects. The evaluation was conducted along two dimensions: data diversity and downstream model performance.
\textcircled{\raisebox{-0.8pt}{3}} We provide both the generated datasets and our implementation code as open-access resources to support broader adoption and reproducibility in the RE community\footnote{\url{https://github.com/abdelkarim-elhajjami/synthline/tree/v0.0.0}}.

The paper is organized as follows: Section 2 reviews the related work relevant to our study. Section 3 introduces the methodology, detailing our PL-based approach and its implementation through Synthline. Section 4 presents our evaluation approach for assessing data diversity and utility. Section 5 discusses the results of our experiments. Section 6 examines threats to validity. Finally, Section 7 concludes the paper and outlines future research directions.
\section{Related Work}
LLMs have motivated numerous studies to explore different approaches for synthetic data generation. The first methods relied on simple context-informed prompting, either in a zero-shot setting using only task specification \cite{b18} or in a few-shot setting, guided by a small number of examples \cite{b13}. Recent works have explored attributed prompting, incorporating explicit attributes to control generation and address issues of low informativeness and redundancy \cite{b34}. Furthermore, multi-step generation approaches decompose the workflow into sequential steps, generating multiple data series to ensure greater diversity \cite{b32}. These studies underscore the importance of controlled generation for both data diversity and quality.

In the RE domain, synthetic data research has primarily focused on data augmentation. Malik et al. \cite{b6} use techniques like back translation, paraphrasing, and RE-specific substitutions (e.g., Noun-Verb or Actor-Action) to improve conflict and duplicate detection in imbalanced datasets. Similarly, Majidzadeh et al. \cite{b7} introduce code-based augmentation methods such as variable renaming and operand swapping, to enhance requirements traceability, yielding improved precision, recall, and F1-score. To the best of our knowledge, Cheng et al. \cite{b8} are the only researchers to employ LLMs (GPT-J) for RE data augmentation; they demonstrated performance gains in classification across three industry case studies.
However, these augmentation approaches have several key limitations. First, they rely on existing datasets, making them ineffective in cases of severe data scarcity. Second, their transformations often produce data that remains close to the original domain and distribution, restricting the potential for genuinely novel data. Third, they typically offer limited control over data attributes such as linguistic, or domain variability. By contrast, our research focuses on generating entirely new synthetic datasets with LLMs, guided by an FM to offer fine-grained controllability.
\section{Overview of the Synthline approach}
This study constitutes the first iteration of a broader design science cycle, where design problems and knowledge questions are iteratively addressed through cycles of improvement design and empirical investigation \cite{b3}.
We chose the PL approach to systematically control and configure the generation of synthetic datasets according to various RE use cases. The PL approach provides a form of systematic customization of a common platform from which multiple variants can be derived \cite{b37}. This paradigm offers two important benefits in our synthetic data generation context. First, it enables systematic customization, ensuring each variant aligns with the specific needs and constraints of its RE use case. Second, it helps improve the quality of the generated data as the platform commonalities can be systematically checked and tested across multiple datasets.
This approach comprises two main phases: \emph{Domain Engineering} and \emph{Application Engineering}. In Domain Engineering, we begin with \textit{domain analysis}, where we define the scope and variability of classification-based RE use cases, resulting in an FM. We then carry out \textit{domain implementation} by developing a configurable generative workflow called \textit{Synthline} as our shared platform. During Application Engineering, we select and configure features from the FM to meet the requirements of a specific RE use case and then use Synthline to generate the corresponding data.
\subsection{Domain Engineering}
\subsubsection{Feature Model}
The first step in the PL approach is domain analysis, which typically results in an FM \cite{b38}. An FM is a widely used notation to describe a PL's variability \cite{b38}. The standard FM notation can represent both concrete and abstract features, marking them as either optional or mandatory. The notation also defines constraints between features to prevent invalid or conflicting configurations. It also supports \emph{or}- and \emph{xor}- decomposition of features, to either allow selection of multiple subfeatures (requiring at least one) from a given feature set, or, restrict selection to exactly one subfeature.

To build the FM, we followed a reactive approach that starts from one or more concrete variants and then generalizes them into an FM. This approach can be done incrementally, gradually refining the model and introducing new features over time, lowering initial costs but potentially taking longer to achieve a stable platform \cite{b40}.
Although our ultimate goal is to cover all RE use cases that can be automated with ML, achieving such broad coverage at once is impractical. Therefore, our incremental reactive approach starts in this study with a subset of RE use cases only. We intend to expand the coverage to additional use cases in future work.

To identify a suitable starting point, we relied on a recent systematic literature review on ML for RE \cite{b12}, which highlighted that classification is the predominant ML task used in the RE domain, significantly surpassing other tasks like clustering and regression. It is important to distinguish between classification as an ML task and requirements classification as an RE use case. Requirements classification, as an RE use case, refers specifically to categorizing requirements into predefined types (e.g., functional vs. non-functional requirements). Classification as an ML task, on the other hand, supports not only requirements classification but also other RE use cases that can be formulated as classification problems, such as identifying requirements specification defects, where requirements are classified based on defect-related labels.
Another key insight from \cite{b12} is that textual requirements artifacts are the primary data source in most RE-related ML studies, compared to other data types such as design documents, domain knowledge, user feedback, or policies and regulations. Based on these findings, we defined the scope of our PL in this study to focus on classification-based RE use cases, with textual requirements as the main artifact. We structured our FM around four core features: \texttt{Generator}, \texttt{Artifact}, \texttt{MLTask}, and \texttt{Output}.

The first feature of the model is \texttt{Generator}, which defines the configuration of the LLM generator. It includes the \texttt{LLM}, \texttt{Temperature}, and \texttt{TopP} features. The \texttt{LLM} subfeature supports two options: GPT-4o and DeepSeek-V3 models. \texttt{Temperature} regulates randomness in the generation process, with lower values producing more deterministic outputs and higher values allowing for greater creativity. \texttt{TopP} enables nucleus sampling, providing fine-grained control over output diversity while maintaining coherence.

The second core feature \texttt{Artifact} defines the characteristics of the generated data through three mandatory subfeatures. The \texttt{Requirement} subfeature, currently the only supported artifact type aligned with our initial scope, contains the core specifications for the requirements content. The \texttt{Domain} specifies the context(s) and \texttt{Language} controls the output language(s).
Within the current \texttt{Requirement} feature, four key aspects are defined. The \texttt{RequirementType} subfeature follows the categorization defined in IEEE 830, including \texttt{ExternalInterfaces} (e.g., user interfaces, hardware interfaces), \texttt{Functions}, \texttt{Performance} requirements, \texttt{LogicalDatabase} specifications, \texttt{DesignConstraints}, and \texttt{SystemAttributes} (e.g., reliability, security, maintainability). The \texttt{SpecificationLevel} allows distinguishing between \texttt{HighLevelSpecification} and \texttt{DetailedSpecification}, enabling generation at different granularity levels. The \texttt{RequirementSource} identifies the stakeholder origin of requirements, such as \texttt{EndUsers}, \texttt{BusinessManagers}, \texttt{DevelopmentTeam}, or \texttt{RegulatoryBodies}. Finally, the \texttt{SpecificationFormat} supports different notation standards including natural language (\texttt{NL}), constrained NL (\texttt{ConstrainedNL}), use cases (\texttt{UseCase}), and user stories (\texttt{UserStory}).

The third core feature \texttt{MLTask} specifies the ML task characteristics that will be applied to the generated data. Within our current scope, we focus on the \texttt{Classification} subfeature, which requires two mandatory components: \texttt{Label} that defines the target classification categories, and \texttt{LabelDescription} that provides detailed specifications of these categories.

The last core feature \texttt{Output} defines the format and size of the generated data. It consists of two distinct subfeatures: \texttt{OutputFormat}, structured as an alternative group that requires choosing between \texttt{CSV} or \texttt{JSON} format, and \texttt{SubsetSize}, enabling control over the volume of data generated in each run.
The complete graphical representation of the FM is provided in the Appendix~\ref{appendix:fm}.
\subsubsection{Synthline Architecture}
\begin{figure}[!ht]
\centering
\includegraphics[width=13cm]{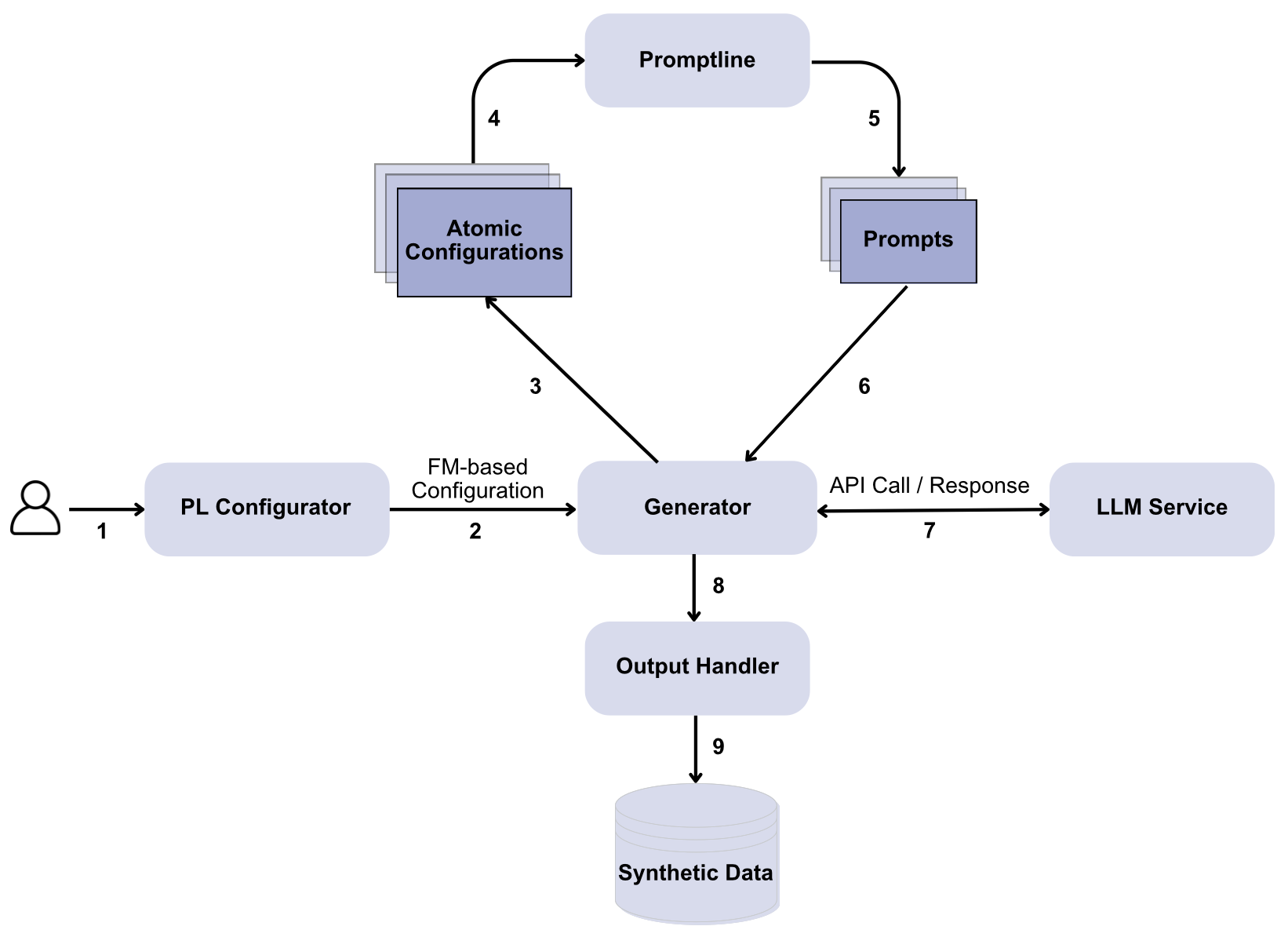}
\caption{High-level architecture of Synthline.} \label{fig0}
\end{figure}

Our application of the design science methodology leads us to design Synthline iteratively. Synthline consists of a PL-based synthetic data generation workflow for RE use cases that can be framed as classification tasks in AI terms.
The architecture of Synthline, illustrated in Figure \ref{fig0}, transforms a configuration of the FM we designed for this purpose into synthetic data through a series of components.

The workflow begins with the PL Configurator, which implements our FM as a graphical interface for feature selection. The interface directly mirrors the hierarchical structure of our FM, organizing controls into four distinct sections corresponding to our core features. For feature selections that require choosing from predefined options, such as \texttt{OutputFormat} between \texttt{CSV} and \texttt{JSON}, the interface provides dropdown menus. For numerical parameters like \texttt{Temperature} or \texttt{SubsetSize}, text input fields are implemented with appropriate validation. This design ensures that users can only create valid configurations while maintaining flexibility in their choices.

The Generator transforms this FM-based configuration into Atomic Configurations, where an Atomic Configuration represents a unique combination of feature values.

The Promptline component converts each Atomic Configuration into a Prompt using a standardized template, based on the structured prompt pattern from [34], which has proven effective:
\begin{quote}
\begin{verbatim}
Generate a requirement that:
1. Is classified as {label} (Description: {label_description})
2. Is of type {requirement_type}
3. Is written in {language}
4. Is for a {domain} system
5. Is from {requirement_source} perspective
6. Follows {specification_format} format
7. Is written at {specification_level} level
Important: Generate ONLY the requirement text. No additional context.
\end{verbatim}
\end{quote}

The Generator manages the distribution of the requested number of samples evenly across all atomic configurations, adjusting for cases where perfect division is not possible. It then interfaces with the LLM Service through asynchronous API calls for text generation. The generated content flows to the Output Handler, which persists the Synthetic Data with its configuration metadata in the specified format.
\subsection{Application Engineering}
\subsubsection{Identification of Requirements Specification Defects}
For demonstrating and evaluating our synthetic data PL approach, we selected the identification of requirements specification defects as our primary use case.

Requirements specification defects present a critical challenge in software development, as they can lead to significant issues in later development stages when problems become increasingly difficult and costly to resolve \cite{b41}. Automated methods to identify these defects, have shown promise in reducing human effort and improving defect detection accuracy \cite{b42,b43,b44}.

The identification of requirements defects can be effectively framed as a classification problem, where each requirement is analyzed to determine the presence of specific types of defects. Several researchers in the RE community have identified different classes of requirements specification defects. Berry and Kamsties \cite{b45} highlighted ambiguity as a primary concern, while other research has emphasized other key defect categories such as non-measurable, non-atomic, and uncertain requirements \cite{b42}.

Table \ref{tab1} details the common requirements specification defects as defined by Fazelnia et al. \cite{b42}. We focus on these defects because we have access to their publicly available dataset\footnote{Dataset available at https://zenodo.org/records/11000349}, in which the requirements are labeled with these six categories.
\begin{table}
\caption{Categories of requirements defects and their definitions \cite{b42}}\label{tab1}
\begin{tabular}{|p{0.25\linewidth}|p{0.75\linewidth}|}
\hline
\textbf{Category} & \textbf{Definition} \\
\hline
Ambiguous Requirements & The requirements specifications are unclear, imprecise, and open to multiple interpretations. \\
\hline
Directive Requirements & Direct the developer to look at additional sources beyond the requirement (e.g., referring to a figure or table). \\
\hline
Non-Measurable Requirements & The requirement is not measurable, or testable, it contain a qualitative value that is unable to be quantified or measured (e.g., “some,” “large,” “long”). \\
\hline
Optional Requirements & Contain terms that may be interpreted in multiple ways and leave the choice of how to implement the requirement to the developer (e.g., "can," "may," "optionally," "as desired," "at last," "either," "eventually," "if appropriate," "in case of," "if necessary"). \\
\hline
Uncertain Requirements & Contain a qualitative value that is insufficiently defined (e.g., "good," "bad," "strong," "easy"), or they may contain under-referenced elements (e.g., "compliant with standards (which ones?)"). \\
\hline
Non-Atomic Requirements & Describe more than one action to be taken and typically have a conjunction, such as "and". \\
\hline
\end{tabular}
\end{table}

The dataset combines requirements from real software systems, including Electronic Health Records Systems from the U.S. Department of Health and Human Services, and high-quality requirements from student projects’ SysRS documents. Two experienced subject matter experts with over 20 and 10 years of experience conducted a rigorous three-stage review process, including grammatical analysis, defect labeling, and final peer review, to ensure high-quality annotations \cite{b42}.

The dataset contains 131 total samples across 6 defect classes, with Ambiguous requirements being the most common (34 samples) and Directive requirements being the least represented (4 samples). This dataset presents two significant limitations: the substantial class imbalance and its relatively small size, which constrains the effective training or fine-tuning of AI models without risking overfitting.
\begin{table}[!ht]
\caption{Distribution of Requirements Specification Defects}\label{tab:defects}
\centering
\begin{tabular}{|l|c|}
\hline
\textbf{Class} & \textbf{\# of samples} \\
\hline
Ambiguous & 34 \\
Directive & 4 \\
Non-Measurable & 18 \\
Optional & 31 \\
Uncertain & 16 \\
Non-Atomic & 28 \\
\hline
\end{tabular}
\end{table}
\subsubsection{Data Generation}
Our synthetic data generation methodology employs each LLM to produce six class-specific subsets aligned with the defect categories defined in Table \ref{tab1}, using their textual definitions as label descriptors. This aligns with the multi-step generation approach that creates multiple data subsets in separate runs, providing more control \cite{b32}.
The configuration settings for our generation process are detailed in Table \ref{tab:configs}.

\begin{table}[!ht]
\caption{Configuration parameters for controlled multi-step generation runs, applied per LLM and defect class in synthetic data subset production.}\label{tab:configs}
\centering
\begin{tabular}{|>{\centering\arraybackslash}m{0.2\linewidth}|
                >{\centering\arraybackslash}m{0.5\linewidth}|
                >{\centering\arraybackslash}m{0.2\linewidth}|}
\hline
\textbf{Feature} & \textbf{Values} & \textbf{\# Atomic Configurations} \\
\hline
Temperature & 1 & 1 \\ 
\hline
TopP & 1 & 1 \\ 
\hline
Requirement Type & External Interfaces, Functions, Performance, Logical Database, Design Constraints, System Attributes & 7 \\ 
\hline
Specification Level & High-Level Specification, Detailed Specification & 2 \\ 
\hline
Requirement Source & End Users, Business Managers, Development Team, Regulatory Bodies & 4 \\ 
\hline
Specification Format & Constrained NL & 1 \\ 
\hline
Language & English & 1 \\ 
\hline
Domain & Healthcare, Restaurant Operations Management & 2 \\ 
\hline
Output Format & CSV & 1 \\ 
\hline
Subset Size & 1120 & 1 \\ 
\hline
\multicolumn{2}{|c|}{\textbf{Total \# Atomic Configurations}} & \textbf{112} \\
\hline
\end{tabular}
\end{table}
We selected GPT-4o and DeepSeek-V3 as our primary LLM options. The inclusion of GPT-4o builds on the demonstrated efficacy of its predecessors in generating high-quality synthetic data \cite{b34},\cite{b50},\cite{b15}. For open-source alternatives, DeepSeek-V3 was selected as the state-of-the-art model as of our experimentation date (5 January 2025), with empirical evaluations indicating it surpasses leading proprietary models, including GPT-4o and Claude 3.5 Sonnet, across most benchmarks \cite{b2}.
Following \cite{b34} and \cite{b50}, a Temperature of 1.0 and TopP of 1.0 were chosen, as these settings have been shown to produce high-quality synthetic data.
The Healthcare domain was selected to directly align with the real dataset’s focus on Electronic Health Records Systems from the U.S. Department of Health and Human Services, while Restaurant Operations Management was chosen as the second domain to reflect the requirements related to restaurant employee registration, menu management, and operational functionalities present in the dataset. Similarly, we chose Constrained NL format and English language to ensure our synthetic requirements mirror the linguistic characteristics of the real dataset.
The CSV output format was chosen for its simplicity and compatibility with the real dataset’s format.
We deliberately aligned our synthetic data features with those of the real dataset, as the evaluation of synthetic models was to be performed on a holdout subset of real data, ensuring a fair assessment of their performance.
Additionally, we included all available subfeatures for requirement types, specification levels, and requirement sources. This comprehensive selection ensures diversity and avoids systematic exclusions that could bias the dataset.

For each LLM (GPT-4o and DeepSeek-V3), we generated 1,120 requirements per defect category, resulting in 6,720 synthetic samples per model and 13,440 samples in total.
\section{Evaluation}
As highlighted in a recent survey \cite{b20}, evaluation methods for synthetic data can be broadly classified into direct and indirect approaches. Direct evaluation focuses on the properties of the generated data itself. A commonly studied aspect is data diversity, which is typically evaluated using vocabulary-based statistics (e.g., vocabulary size and N-gram frequency) or semantic variation metrics such as cosine similarity \cite{b34}. On the other hand, indirect evaluation primarily assesses the utility of synthetic data with respect to the performance of downstream models trained on it, commonly applying a train-on-synthetic, test-on-real paradigm \cite{b34}. In our research, we focus on these two complementary dimensions to comprehensively assess our generated data.

\subsubsection{Diversity Evaluation}
To comprehensively assess the diversity of our synthetic datasets compared to real data, we performed this evaluation after deduplicating\footnote{Deduplication consists of removing identical requirements that were occasionally generated multiple times by the LLMs.} the synthetic datasets (both GPT-4o and DeepSeek). We used several complementary metrics to capture different aspects of textual diversity. We began with two vocabulary-based measurements: absolute vocabulary size, defined as the number of unique terms in each dataset, and normalized vocabulary size per data point, which accounts for varying dataset sizes. These metrics provide a basic indication of lexical richness, where a larger vocabulary suggests greater linguistic variety and potentially richer expression in the generated samples.

For semantic diversity assessment, we computed cosine similarities between text pairs using the embedding of Sentence-BERT \cite{b55}. The Average Pairwise Similarity (APS) measures the overall semantic similarity across all samples in the dataset. A lower APS value indicates greater semantic diversity overall.
The Intra-Class APS, on the other hand, measures the semantic similarity between samples within the same defect class. Lower Intra-Class APS values suggest more diverse examples of the same defect type, indicating less redundancy within classes.

To evaluate phrase-level diversity, we calculated the Inter-sample N-gram Frequency (INGF), which measures repetitive patterns in the generated text. This metric computes the average frequency of n-grams across all samples in a dataset, where lower values indicate less repetition and thus more diverse phrase construction. Together, these metrics provide a multi-faceted view of dataset diversity, from lexical richness to semantic variation and phrase-level uniqueness.

The synthetic datasets’ statistics are summarized in Table~\ref{tab:dataset_stats}, showing the total number of samples and their distribution across defect classes after deduplication.
\begin{table}[!ht]
\caption{Synthetic Datasets’ Statistics After Deduplication.}\label{tab:dataset_stats}
\centering
\begin{tabular}{|l|c|c|}
\hline
\textbf{Defect Class} & \textbf{DeepSeek Samples} & \textbf{GPT-4o Samples} \\
\hline
Ambiguous & 1,007 & 1,117 \\
Directive & 1,015 & 1,120 \\
Non-Atomic & 828 & 1,118 \\
Non-Measurable & 701 & 1,117 \\
Optional & 934 & 1,119 \\
Uncertain & 828 & 1,112 \\
\textbf{Total Samples} & \textbf{5,313} & \textbf{6,703} \\
\hline
\end{tabular}
\end{table}

\subsubsection{Utility Evaluation}
To enable a direct comparison between synthetic and real data performance, we fine-tuned BERT-base-uncased \cite{b35} as the backbone and used the standard cross-entropy loss by default. For hyperparameter selection, we adhered to the recommendations in \cite{b51} and did not use the validation set for model selection. Table \ref{tab:hyperparams} details the hyperparameters used in our experiments.
\begin{table}[!ht]
\caption{Training hyperparameters.}\label{tab:hyperparams}
\centering
\begin{tabular}{|l|c|}
\hline
\textbf{Hyperparameter} & \textbf{Value} \\
\hline
Learning Rate & 5e-5 \\
Batch Size & 32 \\
Training Epochs & 6 \\
Weight Decay & 1e-4 \\
Warmup Ratio & 6\% \\
\hline
\end{tabular}
\end{table}

Our evaluation methodology follows a train-on-synthetic, test-on-real paradigm \cite{b34}. We maintained a holdout test set composed of 30\% of the real data, which remains constant across all experiments to ensure consistent evaluation. The remaining 70\% of real data is used only for training the real-data baseline model. For synthetic data experiments, we trained models on either purely synthetic data or combinations of synthetic and real training data. This approach allows us to directly compare the effectiveness of synthetic data against, and in combination with, real data.

To ensure robust evaluation, we conducted multiple training runs for each configuration, accounting for the inherent variability in model training. We evaluated model performance using macro-averaged precision and recall metrics.
\section{Results}
\subsection{RQ1: How does the diversity of the generated data compare to that of real data?}
As shown in Table \ref{tab:vocab_analysis}, while both GPT-4o and DeepSeek generate datasets with larger absolute vocabularies (4,874 and 3,659 terms respectively) compared to the real dataset (566 terms), this is primarily due to their larger sample sizes. When normalized by sample count, the synthetic datasets demonstrate significantly lower lexical density, with 0.73 and 0.69 words per sample for GPT-4o and DeepSeek respectively, compared to 4.32 words per sample in the real data. This 6.2-6.3× reduction in lexical density suggests that synthetic data tends toward a more constrained vocabulary despite having access to a broader lexical range.
\begin{table}
\caption{Vocabulary size analysis across datasets.}\label{tab:vocab_analysis}
\centering
\begin{tabular}{|l|c|c|}
\hline
\textbf{Dataset} & \textbf{Vocab. Size} & \textbf{Normalized Vocab. Size} \\
\hline
Real Data & 566 & 4.32 \\
DeepSeek & 3,659 & 0.69 \\
GPT-4o & 4,874 & 0.73 \\
\hline
\end{tabular}
\end{table}

\begin{figure}[!ht]
\centering
\includegraphics[width=10cm]{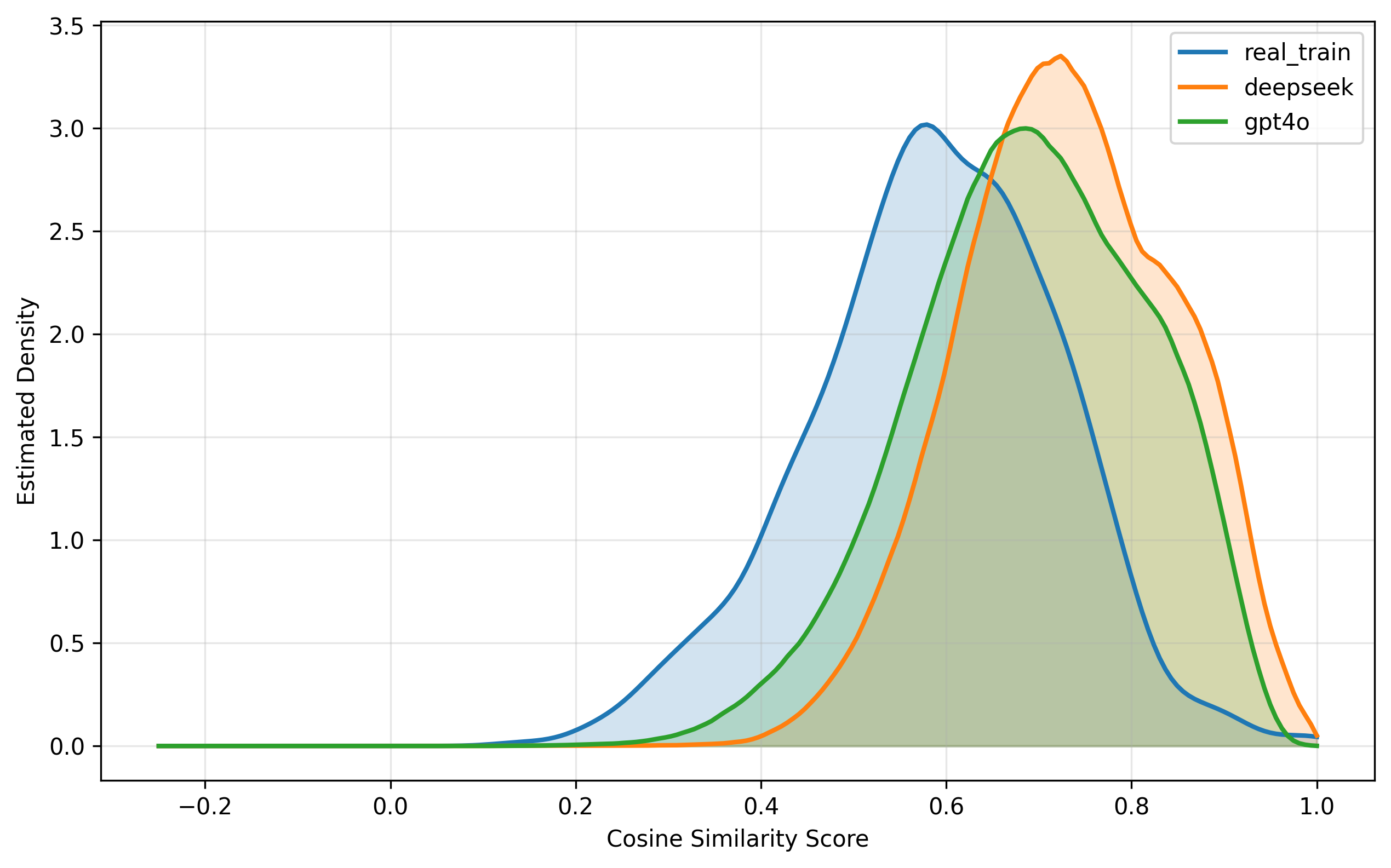}
\caption{Distribution of cosine similarity of text pairs sampled from the same class.} \label{fig1}
\end{figure}

Examining semantic diversity through Table \ref{tab:semantic_metrics}, we observe that synthetic datasets produce more semantically homogeneous content. Both synthetic datasets show higher overall APS values (DeepSeek: 0.649, GPT-4o: 0.612) compared to real data (0.544). This pattern is particularly pronounced in the intra-class analysis, where synthetic data shows notably higher similarity scores (DeepSeek: 0.729, GPT-4o: 0.689) compared to real data (0.587). As illustrated in Figure \ref{fig1}, the distribution of cosine similarities for synthetic data is shifted toward higher values and exhibits a narrower spread.
\begin{table}[!ht]
\caption{Semantic similarity and phrase repetition metrics.}\label{tab:semantic_metrics}
\centering
\begin{tabular}{|l|c|c|c|}
\hline
\textbf{Dataset} & \textbf{APS} & \textbf{Intra-class APS} & \textbf{INGF} \\
\hline
Real Data & 0.544 & 0.587 & 1.214 \\
DeepSeek & 0.649 & 0.729 & 5.086 \\
GPT-4o & 0.612 & 0.689 & 2.472 \\
\hline
\end{tabular}
\end{table}

Table \ref{tab:semantic_metrics} shows that, at the phrase level, synthetic data exhibit significantly higher n-gram repetition, with INGF values 2-4 times higher than real data (DeepSeek: 5.086, GPT-4o: 2.472, vs. Real: 1.214). This increased repetition could be attributed to our choice of Constrained NL Specification format in the generation process, which encourages more standardized phrase construction.

\begin{mdframed}
While synthetic datasets offer broader absolute vocabulary coverage, they show lower diversity than real data across all measured dimensions - exhibiting 6$\times$ lower lexical density, higher semantic similarity (0.612-0.649 vs 0.544), and increased phrase repetition (2-4$\times$ higher INGF values). 
\end{mdframed}
\subsection{RQ2: How does training on synthetic data affect model performance relative to training exclusively on real data?}
We examine model performance across three key training approaches: synthetic-only data, hybrid training combining real and synthetic data, and combined synthetic datasets. First, models trained exclusively on synthetic data show improved performance over the real-data baseline. As shown in Table \ref{tab:training_data_results}, the DeepSeek-based model achieved a 37\% improvement in precision (0.425 vs 0.310), while the GPT-4o-based model showed a 20\% improvement (0.372 vs 0.310). This suggests that synthetic data alone can provide a promising viable training alternative in scenarios where real data is scarce.

\begin{table}[!ht]
\caption{Model performance with different training data compositions. All experiments use identical model architecture, varying only in training data.}\label{tab:training_data_results}
\centering
\begin{tabular}{|l|c|c|}
\hline
\textbf{Training Data} & \textbf{Precision} & \textbf{Recall} \\
\hline
Real data only (baseline) & 0.310 ± 0.050 & 0.256 ± 0.043 \\
DeepSeek only & 0.425 ± 0.083 & 0.334 ± 0.025 \\
GPT-4o only & 0.372 ± 0.060 & 0.329 ± 0.080 \\
GPT-4o + DeepSeek & 0.308 ± 0.037 & 0.297 ± 0.015 \\
Real + DeepSeek & 0.456 ± 0.027 & 0.431 ± 0.042 \\
Real + GPT-4o & 0.575 ± 0.229 & 0.512 ± 0.166 \\
Real + GPT-4o + DeepSeek & 0.469 ± 0.121 & 0.427 ± 0.089 \\
\hline
\end{tabular}
\end{table}

The most notable performance gains, however, emerge from hybrid training approaches combining real and synthetic data. The combination of real and GPT-4o synthetic data yielded the best overall performance, achieving 85\% improvement in precision (0.575 vs. 0.310) compared to the real-data baseline. This substantial improvement suggests strong complementarity between real and synthetic training data.
Interestingly, we observed significant variations in performance stability across different synthetic data sources. Models trained with DeepSeek synthetic data showed more consistent performance (standard deviation: ±0.027 in precision) compared to those trained with GPT-4o data (±0.229). This difference in stability could have important implications for practical applications where performance consistency is important.
A particularly noteworthy finding is that combining multiple synthetic datasets (GPT-4o + DeepSeek) actually degraded performance below that achieved by individual synthetic datasets. This negative synergy persisted even when real data was included in the combination (Real + GPT-4o + DeepSeek: 0.469 vs. Real + GPT-4o: 0.575), suggesting that careful curation of synthetic data sources may be more important than maximizing the volume of synthetic data.

The relatively low performance metrics across all configurations can be attributed to our methodological choice of using recommended hyperparameters \cite{b51} without dataset-specific tuning. While hyperparameter optimization might improve absolute performance, using a fixed configuration across all experiments ensures fair comparisons between different training data compositions, making any relative improvements directly attributable to the data quality.

\begin{mdframed}
Synthetic data can not only match but significantly exceed the performance of real data in training downstream models for classifying requirements defects, with hybrid approaches showing particular promise. However, the success appears highly dependent on the specific combination and curation of synthetic data sources rather than merely increasing data volume.
\end{mdframed}
\section{Threats To Validity}
We identified several threats to validity across the four standard dimensions: internal, external, construct, and conclusion validity. Internally, our models’ performance could be influenced by the specific hyperparameter settings chosen for training. While we followed established recommendations from previous research \cite{b51}, different hyperparameter configurations might yield varying results. Our evaluation methodology's reliance on BERT-base-uncased as the backbone model may introduce bias, as other pre-trained models might perform differently with synthetic data. Additionally, while we followed the structured prompt pattern from \cite{b34}, the synthetic data quality inherently depends on prompt design choices, where even subtle variations in template structure or phrasing could systematically influence the characteristics of generated samples.
Regarding construct validity, our diversity metrics (vocabulary size, APS, INGF) may not fully capture all aspects of diversity in NL generation outputs. While these metrics are widely used in synthetic data evaluation \cite{b20},\cite{b34} and provide useful quantitative measures, they may have inherent limitations in representing the nuanced characteristics that define high-quality diverse text. Our choice to use constrained NL specification format across all synthetic data generation inherently limits linguistic variation and diversity potential, which may affect our assessment of the approach's ability to generate diverse outputs. Moreover, our approach relies on Fazelnia et al.'s \cite{b42} defect category definitions as label descriptions to guide LLM generation, where categories like "Non-Measurable" may have subjective interpretation boundaries, potentially affecting how the LLM interprets and generates examples for each defect type.
Externally, our findings may not generalize beyond the specific context of identifying requirements specification defects to other classification-based RE use cases. The effectiveness of our approach was demonstrated primarily in the healthcare and restaurant operations management domains, and different application domains might present unique challenges not addressed in our current evaluation. Additionally, our results are based on GPT-4o and DeepSeek-V3 models, and the performance characteristics might vary with different LLM versions or alternative models.
Finally, conclusion validity might be affected by the small size of our test set, which could impact the statistical significance of our performance comparisons. Additionally, the variability in model training due to random initialization and stochastic optimization processes introduces potential instability in our results, although we attempted to mitigate this by conducting multiple training runs for each configuration.
\section{Conclusion}
This paper presented Synthline, a synthetic data generation approach that leverages LLMs within a PL strategy to control the generation of synthetic data for classification-based RE use cases. Through an empirical evaluation focused on identifying requirements specification defects, we found that synthetic data, particularly when combined with real data, can significantly enhance downstream classifier performance. Specifically, our experiments showed improvements of up to 85\% in precision and a 2× increase in recall compared to a baseline using only real data. Furthermore, although synthetic datasets exhibited less diversity than real data, the results confirm that carefully curated synthetic samples serve as a viable—and, in many cases, superior—training resource when real data is limited.

In the future, we plan to broaden Synthline's scope to support additional RE use cases and refine generation techniques to further enhance synthetic data quality. We will also develop a more comprehensive evaluation framework that goes beyond diversity and utility, enabling a deeper understanding of how synthetic data contributes to robust, reliable AI solutions in RE.
%
%
%

\newpage
\appendix
\section*{Appendix: Graphical Representation of the FM}
\label{appendix:fm}

\centering
\rotatebox{90}{
    \includegraphics[width=\textheight]{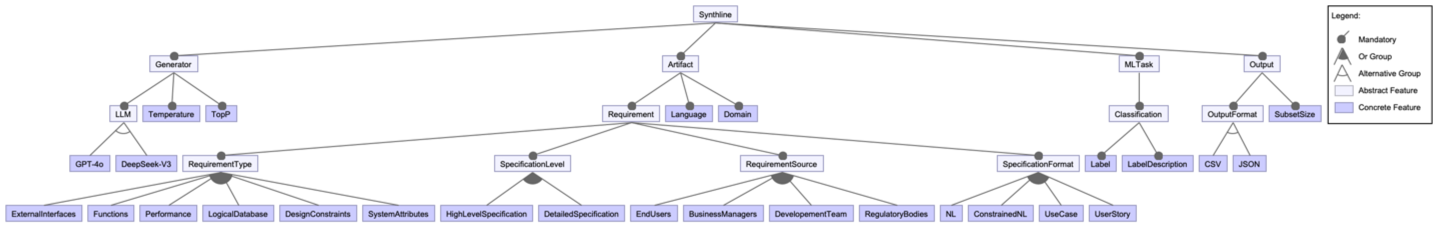}
}


\begin{thebibliography}{30}
\bibitem{b1} X. Franch, C. Palomares, C. Quer, P. Chatzipetrou, and T. Gorschek, "The state-of-practice in requirements specification: an extended interview study at 12 companies", *Requirements Engineering*, pp. 1–33, 2023. doi: 10.1007/s00766-023-00399-7.

\bibitem{b2} DeepSeek-AI, "DeepSeek-V3 Technical Report", arXiv:2412.19437 [cs.CL], 2024.

\bibitem{b3} R. Wieringa, "Design science as nested problem solving", in *Proc. 2009 ACM Int. Conf.*, New York, NY, USA, 2009. doi: 10.1145/1555619.1555630.

\bibitem{b6} G. Malik, M. Cevik, and A. Başar, "Data Augmentation for Conflict and Duplicate Detection in Software Engineering Sentence Pairs", arXiv:2305.09608 [cs.SE], 2023.

\bibitem{b7} A. Majidzadeh, M. Ashtiani, and M. Zakeri-Nasrabadi, "Multi-type requirements traceability prediction by code data augmentation and fine-tuning MS-CodeBERT", *Computer Standards \& Interfaces*, vol. 90, Article 103850, 2024. doi: 10.1016/j.csi.2024.103850.

\bibitem{b8} I. Gräßler, D. Preuß, L. Brandt, and M. Mohr, "Efficient Extraction of Technical Requirements Applying Data Augmentation", in *Proc. 2022 IEEE International Symposium on Systems Engineering (ISSE)*, Vienna, Austria, 2022, pp. 1–8. doi: 10.1109/ISSE54508.2022.10005452.

\bibitem{b9} T. B. Brown, B. Mann, N. Ryder, M. Subbiah, J. Kaplan, P. Dhariwal, A. Neelakantan, P. Shyam, G. Sastry, A. Askell, S. Agarwal, A. Herbert-Voss, G. Krueger, T. Henighan, R. Child, A. Ramesh, and D. M. Ziegler, "Language Models are Few-Shot Learners", arXiv:2005.14165 [cs.CL], 2020.

\bibitem{b10} M. Mosbach, T. Pimentel, S. Ravfogel, D. Klakow, and Y. Elazar, "Few-shot fine-tuning vs. in-context learning: A fair comparison and evaluation", in *Findings of the Association for Computational Linguistics: ACL 2023*, Toronto, Canada, pp. 12284–12314, 2023. Association for Computational Linguistics.

\bibitem{b11} L. Zhao, W. Alhoshan, A. Ferrari, K. J. Letsholo, M. A. Ajagbe, E.-V. Chioasca, and R. T. Batista-Navarro, "Natural Language Processing for Requirements Engineering: A Systematic Mapping Study", *ACM Computing Surveys*, vol. 54, no. 3, Article 55, pp. 1–41, Apr. 2022.

\bibitem{b12} T. Li, X. Zhang, Y. Wang, Q. Zhou, Y. Wang, and F. Dong, "Machine learning for requirements engineering (ML4RE): A systematic literature review complemented by practitioners’ voices from Stack Overflow", *Information and Software Technology*, vol. 172, Article 107477, 2024.

\bibitem{b13} Z. Li, H. Zhu, Z. Lu, and M. Yin, "Synthetic Data Generation with Large Language Models for Text Classification: Potential and Limitations", in *Proc. 2023 Conference on Empirical Methods in Natural Language Processing (EMNLP)*, 2023.

\bibitem{b15} F. Gilardi, M. Alizadeh, and M. Kubli, "ChatGPT outperforms crowd workers for text-annotation tasks", in *Proc. National Academy of Sciences*, vol. 120, no. 30, Article e2305016120, 2023.

\bibitem{b18} J. Ye, J. Gao, Q. Li, H. Xu, J. Feng, Z. Wu, T. Yu, and L. Kong, "ZeroGen: Efficient Zero-shot Learning via Dataset Generation", arXiv:2202.07922 [cs.CL], 2022.

\bibitem{b20} L. Long, R. Wang, R. Xiao, J. Zhao, X. Ding, G. Chen, and H. Wang, "On LLMs-Driven Synthetic Data Generation, Curation, and Evaluation: A Survey", arXiv:2406.15126 [cs.CL], 2024.

\bibitem{b21} A. El-Hajjami, N. Fafin, and C. Salinesi, "Which AI Technique Is Better to Classify Requirements? An Experiment with SVM, LSTM, and ChatGPT", arXiv:2311.11547 [cs.AI], 2024.

\bibitem{b22} J. Yoon, J. Jordon, and M. van der Schaar, "PATE-GAN: Generating Synthetic Data with Differential Privacy Guarantees", in *Proc. International Conference on Learning Representations (ICLR)*, 2019.

\bibitem{b24} S. Abualhaija, F. B. Aydemir, F. Dalpiaz, D. Dell'Anna, A. Ferrari, X. Franch, and D. Fucci, "Replication in Requirements Engineering: the NLP for RE Case", arXiv:2304.10265 [cs.SE], 2024.

\bibitem{b32} Z. Shao, Y. Gong, Y. Shen, M. Huang, N. Duan, and W. Chen, "Synthetic Prompting: Generating Chain-of-Thought Demonstrations for Large Language Models", arXiv:2302.00618 [cs.CL], 2023.

\bibitem{b34} Y. Yu, Y. Zhuang, J. Zhang, Y. Meng, A. Ratner, R. Krishna, J. Shen, and C. Zhang, "Large Language Model as Attributed Training Data Generator: A Tale of Diversity and Bias", in *Proc. Thirty-seventh Conference on Neural Information Processing Systems (NeurIPS) Datasets and Benchmarks Track*, 2023.

\bibitem{b35} J. Devlin, M.-W. Chang, K. Lee, and K. Toutanova, "BERT: Pre-training of Deep Bidirectional Transformers for Language Understanding", in *Proc. 2019 Conference of the North American Chapter of the Association for Computational Linguistics: Human Language Technologies, Volume 1 (Long and Short Papers)*, J. Burstein, C. Doran, and T. Solorio, Eds., Minneapolis, Minnesota: Association for Computational Linguistics, 2019, pp. 4171–4186. doi: 10.18653/v1/N19-1423.

\bibitem{b37} K. Pohl, G. Böckle, and F. Van Der Linden, *Software Product Line Engineering: Foundations, Principles, and Techniques*, vol. 1, Berlin, Germany: Springer, 2005.

\bibitem{b38} S. Apel, D. Batory, C. Kästner, and G. Saake, *Feature-Oriented Software Product Lines: Concepts and Implementation*, Springer, Berlin, Heidelberg, 2013.

\bibitem{b40} L. Northrop, "Software product lines essentials", *Pittsburgh: SEI Carnegie Mellon University*, 2008.

\bibitem{b41} S. Lauesen and O. Vinter, “Preventing Requirement Defects: An Experiment in Process Improvement”, in Requirements Engineering Journal, vol. 6, no. 1, pp. 37-50, 2001

\bibitem{b42} M. Fazelnia, V. Koscinski, S. Herzog, and M. Mirakhorli, “Lessons from the Use of Natural Language Inference (NLI) in Requirements Engineering Tasks”, in Proc. 32nd IEEE International Requirements Engineering Conference (RE), 2024.

\bibitem{b43} S. Goyal et al., “SDP-BB: A Software Defect Prediction Model Using BiLSTM and BERT-Based Semantic Features”, in IEEE Transactions on Software Engineering, 2022.

\bibitem{b44} A. Dautovic, R. Plösch, and M. Saft, “Automated Quality Defect Detection in Software Development Documents”, in Proc. 5th International Conference on Software Quality Management, 2011.

\bibitem{b45} D. M. Berry and E. Kamsties, “Ambiguity in Requirements Specification”, in Requirements Engineering: State of the Practice, M. Leite and J. Doorn, Eds., Boston, MA: Springer US, 2004, pp. 7-44.

\bibitem{b50} B. Peng, C. Li, P. He, M. Galley, and J. Gao, "Instruction Tuning with GPT-4", arXiv:2304.03277 [cs.CL], 2023.

\bibitem{b51} E. Perez, D. Kiela, and K. Cho, "True Few-Shot Learning with Language Models", in *Proc. Advances in Neural Information Processing Systems*, A. Beygelzimer, Y. Dauphin, P. Liang, and J. Wortman Vaughan, Eds., 2021.

\bibitem{b55} N. Reimers and I. Gurevych, "Sentence-BERT: Sentence Embeddings using Siamese BERT-Networks", in *Proc. 2019 Conference on Empirical Methods in Natural Language Processing and the 9th International Joint Conference on Natural Language Processing (EMNLP-IJCNLP)*, K. Inui, J. Jiang, V. Ng, and X. Wan, Eds., Hong Kong, China: Association for Computational Linguistics, 2019, pp. 3982–3992. doi: 10.18653/v1/D19-1410.
\end{thebibliography}
\end{document}